\documentclass[global]{journal}

\smartqed

\usepackage{graphicx}
\usepackage{mathptmx}
\usepackage{ispo2}

\journalname{xxxxxxx}

\begin{document}

\sloppy

\title{Turing Impossibility Properties for Stack Machine Programming}

\author{J.A. Bergstra \and C.A. Middelburg}

\institute{Informatics Institute, Faculty of Science,
           University of Amsterdam, \\
           Science Park~904, 1098~XH Amsterdam, the Netherlands \\
           \email{J.A.Bergstra@uva.nl, C.A.Middelburg@uva.nl}}

\date{}

\maketitle

\begin{abstract}
The strong, intermediate, and weak Turing impossibility properties are
introduced. Some facts concerning Turing impossibility for stack
machine programming are trivially adapted from previous work.
Several intriguing questions are raised about the Turing impossibility
properties concerning different method interfaces for stack machine
programming.
\end{abstract}

\section{Introduction}
\label{sect-intro}

The work presented in this paper constitutes a minor adaptation to a 
simplified setting, and a corresponding reformulation of the content of our \cite{BM09k}.
We refer to that paper for further technical
explanations of the formalism used below, for the justification of terminology,
as well as for more information
concerning connections with previous work.%
\footnote{In \cite{BM09k} the focus is on modeling Turing machine computation, while in this
paper the focus is on stack machines. In addition \cite{BM09k} explains the
semantics of instruction sequences via thread algebra (see \cite{BM07a}) and in this paper we
will use an operational semantics instead.}
  By highlighting results from
\cite{BM09k} from a  different perspective their relevance for
understanding the methodological impact of the well-known recursive
unsolvability of the halting problem, in which we firmly believe,
becomes more apparent. Like \cite{BM09k} this paper focuses on the off-line
halting problem, which unlike the on-line halting problem analyzed in \cite{BP04a}
need not always give way to a diagonal argument.

This paper concerns  an investigation of issues
relating to the halting problem perceived in terms of instruction sequences.
Positioning Turing's result of~\cite{Tur37a} regarding the recursive unsolvability of the halting
problem  as a result about programs rather than
machines, and taking instruction sequences as programs, we analyse the
autosolvability requirement that a program of a certain kind must solve
the halting problem for all programs of that kind.

Below we will use the term execution both in connection with instructions and in 
connection with instruction sequences. This is not entirely consistent with \cite{Bergstra2011a}
where execution is given a rather confined meaning, involving the use of real computing devices.
Here instruction sequences are mathematical objects and their execution, by necessity is merely 
a mathematical or logical model for the (real) putting into effect of (physical representations of)
instruction sequences.%
\footnote{In fact execution as used in this paper corresponds to 
``directly putting into effect'' as used in \cite{Bergstra2011a}. Instructions are said to be executed as well, 
or alternatively instructions are said to be issued, 
thus following a common terminology in computer architecture.}

The paper follows the organization of \cite{BM09k}, beginning with
 a survey of the instruction sequence notation that will be used used in this paper
(Section~\ref{sect-PGLBbt}).
Next, we introduce services and a composition operator for services
families (Section~\ref{sect-SF}). In Section~\ref{sect-TSI} an operational semantics is provided
for instruction sequences under execution in a context of service families.
In Section \ref{sect-a-r-ops}  following \cite{BM09k}, 
we add two operators, named $\sfapply$, and  $\sfreply$, 
that are
related to the processing of instructions by a service family.
Then, as in \cite{BM09k} we propose to comply with conventions that exclude the use of
terms that are not really intended to denote anything
(Sections~\ref{sect-RUC}).
Thereafter, we introduce the concept of a functional unit and related
concepts (Section~\ref{sect-func-unit}).
Then, we define autosolvability and related notions in terms of
functional units related to stack machines
(Section~\ref{sect-func-unit-sbs}). In Section \ref{sect-stack-methods} we specify a familiar menu
of method names and operations on stacks.  In Section \ref{sect-t-i-properties} 
we introduce the strong, intermediate,
and weak Turing impossibility properties for programming environments.
equipped with a given and fixed way to encode instruction sequences into
functional unit states.
After that, we  give
positive and negative results concerning the autosolvability of the
halting problem (Section~\ref{sect-autosolvability}).  In Section \ref{sect-open-problems}
we provide a number of questions concerning Turing impossibility for
stack machine programming.
Finally, we make some concluding remarks (Section~\ref{sect-concl}).

\section{PGLB with Boolean Termination}
\label{sect-PGLBbt}

In this section, we introduce the program notation \PGLBbt\ (PGLB with
Boolean termination).
In~\cite{BL02a}, a hierarchy of program notations rooted in program
algebra is presented.
One of the program notations that belong to this hierarchy is \PGLB\
(ProGramming Language B).
This program notation is close to existing assembly languages and has
relative jump instructions.
\PGLBbt\ is \PGLB\ extended with two termination instructions that allow
for the execution  of an instruction sequence   to yield a Boolean value at
termination.
The extension makes it possible to deal naturally with instruction
sequences that implement some test, which is relevant throughout the
paper.

In \PGLBbt, it is assumed that a fixed but arbitrary non-empty finite
set $\BInstr$ of \emph{basic instructions} has been given.
The intuition is that the issuing of a basic instruction in most
instances effects the modification of a state and in all instances 
produces a reply at its completion.
The possible replies are $\True$ (standing for true) and $\False$
(standing for false), and the actual reply is in most instances
state-dependent.
Therefore, successive executions of the same basic instruction may
produce different replies.

\PGLBbt\ has the following primitive instructions:
\begin{itemize}
\item
for each $a \in \BInstr$, a \emph{plain basic instruction} $a$;
\item
for each $a \in \BInstr$, a \emph{positive test instruction} $\ptst{a}$;
\item
for each $a \in \BInstr$, a \emph{negative test instruction} $\ntst{a}$;
\item
for each $l \in \Nat$, a \emph{forward jump instruction}
$\fjmp{l}$;
\item
for each $l \in \Nat$, a \emph{backward jump instruction}
$\bjmp{l}$;
\item
a \emph{plain termination instruction} $\halt$;
\item
a \emph{positive termination instruction} $\haltP$;
\item
a \emph{negative termination instruction} $\haltN$.
\end{itemize}
\PGLBbt\ instruction sequences have the form
$u_1 \conc \ldots \conc u_k$, where $u_1,\ldots,u_k$ are primitive
instructions of \PGLBbt.

In the process of executing a \PGLBbt\ instruction sequence, 
these primitive instructions have the following effects:
\begin{itemize}
\item
the effect of a positive test instruction $\ptst{a}$ is that basic
instruction $a$ is executed and the execution proceeds with the next
primitive instruction if $\True$ is produced and otherwise the next
primitive instruction is skipped and the execution proceeds with the
primitive instruction following the skipped one -- if there is no
primitive instruction to proceed with, deadlock occurs;
\item
the effect of a negative test instruction $\ntst{a}$ is the same as the
effect of $\ptst{a}$, but with the role of the value produced reversed;
\item
the effect of a plain basic instruction $a$ is the same as the effect of
$\ptst{a}$, but a run always proceeds as if $\True$ is produced;
\item
the effect of a forward jump instruction $\fjmp{l}$ is that the execution
proceeds with the $l^\mathrm{th}$ next primitive instruction -- if $l$
equals $0$ or there is no primitive instructions to proceed with,
deadlock occurs;
\item
the effect of a backward jump instruction $\bjmp{l}$ is that the execution
proceeds with the $l^\mathrm{th}$ previous primitive instruction -- if
$l$ equals $0$ or there is no primitive instruction to proceed with,
deadlock occurs;
\item
the effect of the plain termination instruction $\halt$ is that
the execution terminates and in doing so does not deliver a value;
\item
the effect of the positive termination instruction $\haltP$ is that
the execution terminates and in doing so delivers the Boolean value $\True$;
\item
the effect of the negative termination instruction $\haltN$ is that
the execution terminates and in doing so delivers the Boolean value
$\False$.
\end{itemize}

A simple example of a \PGLBbt\ instruction sequence is
\begin{ldispl}
\ptst{a} \conc \fjmp{2} \conc \bjmp{2} \conc b \conc \haltP\;.
\end{ldispl}%
When executing this instruction sequence, first the basic instruction
$a$ is issued repeatedly until its execution produces the reply
$\True$, next the basic instruction $b$ is executed, and after that
the run terminates with delivery of the value $\True$.

From Section~\ref{sect-func-unit}, we will use a restricted version of
\PGLBbt\ called \PGLBsbt\ (\PGLB\ with strict Boolean termination).
The primitive instructions of \PGLBsbt\ are the primitive instructions
of \PGLBbt\ with the exception of the plain termination instruction.
Thus, \PGLBsbt\ instruction sequences are \PGLBbt\ instruction sequences
in which the plain termination instruction does not occur.

We will write $\Inseq$ to denote the set of \PGLBbt\ instruction sequences
below. We will view this set of instruction sequences as a sort in a many-sorted algebra
for which the sort name $\Inseq$ will be used.  Further each \PGLBbt\ instruction sequence
is used as a constant of sort $\Inseq$ denoting itself.%
\footnote{This treatment of the sort $\Inseq$ is a shortcut of the presentation of \cite{BM09k}
and \cite{BL02a} where the sort of threads is used as a behavioral abstraction of
instruction sequences.}

\section{Services and Service Families}
\label{sect-SF}

In this section, we introduce service families and a composition
operator for service families.
We start by introducing services.

It is assumed that a fixed but arbitrary non-empty finite set $\Meth$
of \emph{methods} has been given.
A service is able to process certain methods.
The processing of a method may involve a change of the service.
At completion of the processing of a method, the service produces a
reply value.
The set $\Replies$ of \emph{reply values} is the set
$\set{\True,\False,\Div}$.
The reply value $\Div$ stands for divergent.

For example, a service may be able to process methods for pushing a
natural number on a stack ($\push{n}$), testing whether the top of the
stack equals a natural number ($\topeq{n}$), and popping the top element
from the stack ($\pop$).
Execution of a pushing method or a popping method changes the service,
because it changes the stack with which it deals, and produces the reply
value $\True$ if no stack overflow or stack underflow occurs and
$\False$ otherwise.
Execution of a testing method does not change the service, because it
does not changes the stack with which it deals, and produces the reply
value $\True$ if the test succeeds and $\False$ otherwise.
Attempted processing of a method that the service is not able to process
changes the service into one that is not able to process any method and
produces the reply $\Div$.

In \SFA, the algebraic theory of service families introduced below, the
following is assumed with respect to services:
\begin{itemize}
\item
a set $\Services$ of services has been given together with:
\begin{itemize}
\item
for each $m \in \Meth$,
a total function $\funct{\effect{m}}{\Services}{\Services}$;
\item
for each $m \in \Meth$,
a total function $\funct{\sreply{m}}{\Services}{\Replies}$;
\end{itemize}
satisfying the condition that there exists a unique $S \in \Services$
with $\effect{m}(S) = S$ and $\sreply{m}(S) = \Div$ for all
$m \in \Meth$;

When dealing with examples and applications we will assume that a
sufficiently large collection of services is available and
that a name is known for each of those. In addition for each name all equations
that determine the graphs of $\effect{m}$ and $\sreply{m}$ are available.
\item
a signature $\Sig{\Services}$ has been given that includes the following
sort:
\begin{itemize}
\item
the sort $\Serv$ of \emph{services};
\end{itemize}
and the following constant and operators:
\begin{itemize}
\item
the \emph{empty service} constant $\const{\emptyserv}{\Serv}$;
\item
for each $m \in \Meth$,
the \emph{derived service} operator
$\funct{\derive{m}}{\Serv}{\Serv}$;
\end{itemize}
\item
$\Services$ and $\Sig{\Services}$ are such that:
\begin{itemize}
\item
each service in $\Services$ can be denoted by a closed term of sort
$\Serv$;
\item
the constant $\emptyserv$ denotes the unique $S \in \Services$ such
that $\effect{m}(S) = S$ and $\sreply{m}(S) = \Div$ for all
$m \in \Meth$;
\item
if closed term $t$ denotes service $S$, then $\derive{m}(t)$ denotes
service $\effect{m}(S)$.
\end{itemize}
\end{itemize}

When a request is made to service $S$ to process method $m$:
\begin{itemize}
\item
if $\sreply{m}(S) \neq \Div$, then $S$ processes $m$, produces the reply
$\sreply{m}(S)$, and next proceeds as $\effect{m}(S)$;
\item
if $\sreply{m}(S) = \Div$, then $S$ is not able to process method $m$
and proceeds as $\emptyserv$.
\end{itemize}
The empty service $\emptyserv$ is the unique service that is unable to
process any method.

It is also assumed that a fixed but arbitrary non-empty finite set
$\Foci$ of \emph{foci} has been given.
Foci play the role of names of services in the service family offered by
an execution architecture.
A service family is a set of named services where each name occurs only
once.

\SFA\ has the sorts, constants and operators in $\Sig{\Services}$
and in addition the following sort:
\begin{itemize}
\item
the sort $\ServFam$ of \emph{service families};
\end{itemize}
and the following constant and operators:
\begin{itemize}
\item
the \emph{empty service family} constant $\const{\emptysf}{\ServFam}$;
\item
for each $f \in \Foci$, the unary \emph{singleton service family}
operator $\funct{\mathop{f{.}} \ph}{\Serv}{\ServFam}$;
\item
the binary \emph{service family composition} operator
$\funct{\ph \sfcomp \ph}{\ServFam \x \ServFam}{\ServFam}$;
\item
for each $F \subseteq \Foci$, the unary \emph{encapsulation} operator
$\funct{\encap{F}}{\ServFam}{\ServFam}$.
\end{itemize}
We assume that there is a countably infinite set of variables of sort
$\ServFam$ which includes $u,v,w$.
Terms are built as usual in the many-sorted case
(see e.g.~\cite{Wir90a,ST99a}).
We use prefix notation for the singleton service family operators and
infix nota\-tion for the service family composition operator.

The service family denoted by $\emptysf$ is the empty service family.
The service family denoted by a closed term of the form $f.H$ consists
of one named service only, the service concerned is the service denoted
by $H$, and the name of this service is $f$.
The service family denoted by a closed term of the form $C \sfcomp D$
consists of all named services that belong to either the service family
denoted by $C$ or the service family denoted by $D$.
In the case where a named service from the service family denoted by $C$
and a named service from the service family denoted by $D$ have the same
name, they collapse to an empty service with the name concerned.
The service family denoted by a closed term of the form $\encap{F}(C)$
consists of all named services with a name not in $F$ that belong to the
service family denoted by $C$.
Thus, the service families denoted by closed terms of the forms $f.H$
and $\encap{\set{f}}(C)$ do not collapse to an empty service in service
family composition.

Using the singleton service family operators and the service family
composition operator, any finite number of possibly identical services
can be brought together in a service family provided that the services
concerned are given different names.

The empty service family constant and the encapsulation operators are
primarily meant to axiomatize the operators that are introduced in
Section~\ref{sect-TSI}.

The axioms of \SFA\ are given in Table~\ref{axioms-SFA}.%
\begin{table}[!t]
\caption{Axioms of \SFA}
\label{axioms-SFA}
\begin{eqntbl}
\begin{axcol}
u \sfcomp \emptysf = u                                 & \axiom{SFC1} \\
u \sfcomp v = v \sfcomp u                              & \axiom{SFC2} \\
(u \sfcomp v) \sfcomp w = u \sfcomp (v \sfcomp w)      & \axiom{SFC3} \\
f.H \sfcomp f.H' = f.\emptyserv                        & \axiom{SFC4}
\end{axcol}
\qquad
\begin{saxcol}
\encap{F}(\emptysf) = \emptysf                       & & \axiom{SFE1} \\
\encap{F}(f.H) = \emptysf & \mif f \in F               & \axiom{SFE2} \\
\encap{F}(f.H) = f.H      & \mif f \notin F            & \axiom{SFE3} \\
\encap{F}(u \sfcomp v) =
\encap{F}(u) \sfcomp \encap{F}(v)                    & & \axiom{SFE4}
\end{saxcol}
\end{eqntbl}
\end{table}
In this table, $f$ stands for an arbitrary focus from $\Foci$ and $H$
and $H'$ stand for arbitrary closed terms of sort $\Serv$.
The axioms of \SFA\ simply formalize the informal explanation given above.

The \emph{foci} operation $\foci$ defined by the
equations in Table~\ref{eqns-foci}
(for foci $f \in \Foci$ and terms $H$ of sort $\Serv$) provides the collection of
foci that occur within a service family. Knowledge of this collection
plays a role when defining
the operational semantics of instruction sequences acting on a service family.%
\begin{table}[!t]
\caption{Defining equations for the foci operation}
\label{eqns-foci}
\begin{eqntbl}
\begin{eqncol}
\foci(\emptysf) = \emptyset                                           \\
\foci(f.H) = \set{f}                                                  \\
\foci(u \sfcomp v) = \foci(u) \union \foci(v)
\end{eqncol}
\end{eqntbl}
\end{table}
The operation $\foci$ gives, for each service family, the set of all
foci that serve as names of named services belonging to the service
family.

Given a  service family $C$, if  $f \not\in \foci(C)$ then $C$ can be written as
$\encap{\{f\}}(C^{\prime})$, and if $f \in \foci(C)$ then $C$ can be written as
$f.H \union \encap{\{f\}}(C^{\prime})$ for a suitable service $H$ and
an appropriate service family $C^{\prime}$.

\section{Operational semantics}
\label{sect-TSI}
For the set $ \BInstr$ of basic instructions, we take the set
$\set{f.m \where f \in \Foci, m \in \Meth}$.
Let $1 \leq i \leq k$, and let $p = u_1 \conc \ldots \conc u_k$ be a  \PGLBbt\ instruction sequence, 
with basic instructions in $ \BInstr$,
and for that reason a closed $\Inseq$ term and let $C$ denote a service family and at the same time
a closed $\ServFam$ term. Then a triple $\config{i,p,C}$ can be read
as the configuration consisting of  $p$ acting on service family $C$ with program
counter at value $i$ when $p$ is executed. Configurations are computational states
but we will only use the term state for services and service families, and speak of a configuration if
the instruction sequence is included as well as positional information about the instruction which is next to be
issued, that is a program counter.

From a non-terminal configuration $\config{i,p,C}$,
subsequent computational steps start with issuing the $i^\mathrm{th}$ primitive instruction, i.e.\
$u_i$. By default, a run starts at the first primitive instruction. For technical reasons
configurations with $i=0$ or $i > k$ will be considered as well.

The operational semantics describes how a configuration can develop step by step
into other configurations. Terminal configurations are configurations that satisfy any of the following conditions:
\begin{itemize}
\item $u_i = \halt$, or $u_i = \haltP$, or $u_i = \haltN$, or
\item $i=0$ or $i > 0$, or
\item $u_i \equiv f.m$, or $u_i \equiv +f.m$, or $u_i \equiv -f.m$ for a focus $f$ such that $f \not\in \foci(C)$.
\end{itemize}

If $u_i = \halt$, or $u_i = \haltP$, or $u_i = \haltN$, then the configuration is correctly terminating.
In all other cases the terminating configuration specifies an erroneous state indicating incorrect termination.%
\footnote{Incorrect termination can be understood to represent the occurrence of an error during a computation. 
For instance execution of the instruction sequence $\#1;\backslash \#5;\halt$ will lead to an 
error after the first instruction has been executed and for that reason the backward jump in the second instruction
constitutes a fault in the instruction sequence.}

The sequence of steps  from a configuration is called a computation. 
Each step involves either the execution of a jump or the application of a method to a service.
The service involved in the processing of a method is the service whose
name is the focus of the basic instruction in question. After proceeding 0 or more steps a 
computation can but need not end in a terminal configuration. If it ends in a terminal configuration the
computation is said to converge, otherwise it proceeds forever and it is
said to diverge. If the terminal configuration is  correctly terminating, the computation is 
said to be successful, otherwise the terminal configuration is incorrectly terminating and the 
computation is said to be unsuccessful.

Computation steps for configurations are generated by the following four rules:%
\footnote
{As usual, we write $i \monus j$ for the monus of $i$ and $j$, i.e.\
 $i \monus j = i - j$ if $i \geq j$ and $i \monus j = 0$ otherwise.
}

\begin{enumerate}
\item
$\Rule{u_i \equiv \# k}{\config{i,p,C} \step{\texttt{fw-jmp}} \config{i+k,p,C}}$

\item
$\Rule{u_i \equiv \backslash \# k}{
\config{i,p,C} \step{\texttt{bw-jmp}} \config{i \monus k,p,C}}$%

\item
$\Rule{u_i \equiv f.m 
\vee (u_i \equiv +f.m \wedge  \sreply{m}(H) = \True) 
\vee (u_i \equiv -f.m \wedge  \sreply{m}(H) = \False)}{
\config{i,p,f.H \sfcomp \encap{\set{f}}(C)}
\step{\texttt{b-act}}
\config{i+1,p,f.\derive{m}H \sfcomp \encap{\set{f}}(C))}}$

\item
$\Rule{(u_i \equiv +f.m \wedge \sreply{m}(H) = \False)
\vee (u_i \equiv -f.m \wedge  \sreply{m}(H) = \True)} {
\config{i,p,f.H \sfcomp \encap{\set{f}}(C)} \step{\texttt{b-act}} \config{i+2,p,f.\derive{m}H \sfcomp \encap{\set{f}}(C))}
}$
\end{enumerate}

An instruction sequence may interact with the named services from the service family
offered by an execution architecture.
That is, during its executed an instruction sequence
may issue a basic instruction for the purpose of
requesting a named service to process a method and to return a reply
value at completion of the processing of the method.

\section{Apply and reply operators}
\label{sect-a-r-ops}
In this section, we combine the sort $\Inseq$ with the sort $\ServFam$ and extend the
combination with two operators, called apply operator and reply operator respectively, 
that relate to this kind of interaction between instruction sequences and services.

The reply operator is concerned with the effects of service families on
the Boolean values that computations possibly deliver at their termination.
The reply operator does not always produce Boolean values: it produces
special values in cases where no Boolean value is delivered at
termination or no termination takes place. The apply operator determines the
successive effect that basic instructions issued during a terminating execution 
have on a service family. The apply operator is made total by stipulating that it 
produces the empty service family in the case of  diverging computations.

Both operators mentioned above are concerned with the processing of
methods by services from a service family in pursuance of basic instructions
issued when an instruction sequence is executed.

We will use  in addition the following sort:
\begin{itemize}
\item
the sort $\Repl$ of \emph{replies};
\end{itemize}
and the following constants and operators:
\begin{itemize}
\item
the \emph{reply} constants $\const{\True,\False,\Div,\Mea}{\Repl}$;
\item
the binary \emph{apply} operator
$\funct{\ph \sfapply \ph}{\Inseq \x \ServFam}{\ServFam}$;
\item
the binary \emph{reply} operator
$\funct{\ph \sfreply \ph}{\Inseq \x \ServFam}{\Repl}$.
\end{itemize}
We use infix notation for the apply and reply operators.

The service family denoted by a closed term of the form $p \sfapply C$
is the service family that results from processing
the method of each basic instruction issued by the instruction sequence $p$
by the service in the service family denoted by $C$ with the focus
of the basic instruction as its name if such a service exists.

The value denoted by  $p \sfreply C$ is the
Boolean value serving as the flag of the termination instruction at which
computation starting from the initial configuration $\config{1,p,C}$ comes to a halt  if that 
computation terminates
correctly and in addition this termination instruction carries a Boolean value. 

The value $\Mea$ (standing for meaningless) is
yielded if the computation
terminates correctly  ending with the program counter at a termination instruction not carrying a 
Boolean value, and the result is the
value $\Div$ (standing for divergent) if the computation does not correctly terminate.
Formally the connection between computations and the apply and reply operators is as follows (again
assuming that $k$ is the number of instructions in $p= u_1 \conc \ldots \conc u_k$):
\begin{itemize}
\item if $\config{1,p,C}$ produces a divergent computation then $p \sfapply C = \emptysf$
and $p \sfreply C = \Div$.
\item if $\config{1,p,C}$ produces an incorrectly terminating computation, say in some configuration
 $\config{i,p,D}$ that satisfies one of these five conditions:
$i=0$, or $i > k$, or $u_i \equiv f.m$,  or $u_i \equiv+f.m$, or  $u_i \equiv -f.m$ for some basic instruction 
$f.m$ (for which $f \not\in \foci(D)$ must necessarily hold), 
then $p \sfapply C = \emptysf$ and $p \sfreply C = \Div$. 
\item if $\config{1,p,C}$ produces a correctly terminating computation, say ending in a configuration,
 $\config{i,p,D}$ such that $1 \leq i \leq k$, and either $u_i \equiv \halt$, or  $u_i \equiv \haltP$, 
or $u_i \equiv \haltN$, then $p \sfapply C = D$. Further in this case: if $u_i \equiv \halt$ then
$p \sfreply C = \Mea$, if $u_i \equiv \haltP$ then $p \sfreply C = \True$, and if $u_i \equiv \haltN$
then $p \sfreply C = \False$. 
\end{itemize}

We write $p \cvg u$ iff $p \sfreply u = \True$ or $p \sfreply u = \False$ or
$p \sfreply u = \Mea$. We write $p \cvgb u$ iff $p \sfreply u = \True$ or $p \sfreply u = \False$.

\section{Relevant Use Conventions}
\label{sect-RUC}

In the setting of service families, sets of foci play the role of
interfaces.
The set of all foci that serve as names of named services in a service
family is regarded as the interface of that service family.
There are cases in which processing does not terminate or, even worse
(because it is statically detectable), interfaces of services families
do not match.
In the case of non-termination, there is nothing that we intend to
denote by a term of the form $p \sfapply C$ or $p \sfreply C$.
In the case of non-matching services families, there is nothing that we
intend to denote by a term of the form $C \sfcomp D$.
Moreover, in the case of termination without a Boolean reply, there is
nothing that we intend to denote by a term of the form $p \sfreply C$.

We propose to comply with the following \emph{relevant use conventions}:
\begin{itemize}
\item
$p \sfapply C$ is only used if it is known that $p \cvg C$;
\item
$p \sfreply C$ is only used if it is known  that $p \cvgb C$;%
\footnote{If it turns out  that in some case $p \sfreply C = \False$ a 
failure has occurred because by using If $p \sfreply C$ the belief
is implicitly assumed that $p \cvgb C.$ A plausible cause for that
state of affairs is an instruction sequencing fault. That is a mismatch 
between instruction sequencer intentions and the operational semantics 
of the instruction sequence that was constructed, for instance if at some place
$\halt$
was written where $\haltP$ was meant. Another plausible cause 
is that a mistake was made concerning the choice 
which instruction sequence from a library of given ones to execute.}

\item
$C \sfcomp D$ is only used if it is known that
$\foci(C) \inter \foci(D) = \emptyset$.
\end{itemize}

The condition found in the first convention is justified by the fact
that $x \sfapply u = \emptysf$ if $x \dvg u$.
We do not have $x \sfapply u = \emptysf$ only if $x \dvg u$.
For instance, $\haltP \sfapply \emptysf = \emptysf$ whereas
$\haltP \cvg \emptysf$.
Similar remarks apply to the condition found in the second convention.

The idea of relevant use conventions is taken from~\cite{BM09g}, where
it plays a central role in an account of the way in which mathematicians
usually deal with division by zero in mathematical texts.
In the sequel, we will comply with the relevant use conventions
described above.

\section{Functional Units}
\label{sect-func-unit}

In this section, we introduce the concept of a functional unit and
related concepts.

It is assumed that a non-empty finite or countably infinite set $\FUS$
of \emph{states} has been given.
As before, it is assumed that a non-empty finite set $\MN$ of methods
has been given.
However, in the setting of functional units, methods serve as names of
operations on a state space.
For that reason, the members of $\MN$ will henceforth be called
\emph{method names}.

A \emph{method operation} on $\FUS$ is a total function from $\FUS$ to
$\Bool \x \FUS$.
A \emph{partial method operation} on $\FUS$ is a partial function from
$\FUS$ to $\Bool \x \FUS$.
We write $\MO(\FUS)$ for the set of all method operations on $\FUS$.
We write $M^r$ and $M^e$, where $M \in \MO(\FUS)$, for the unique
functions $\funct{R}{\FUS}{\Bool}$ and $\funct{E}{\FUS}{\FUS}$,
respectively, such that $M(s) = \tup{R(s),E(s)}$ for all $s \in \FUS$.

A \emph{functional unit} for $\FUS$ is a finite subset $\cH$ of
$\MN \x \MO(\FUS)$ such that $\tup{m,M} \in \cH$ and
$\tup{m,M'} \in \cH$ implies $M = M'$.
We write $\FU(\FUS)$ for the set of all functional units for $\FUS$.
We write $\IF(\cH)$, where $\cH \in \FU(\FUS)$, for the set
$\set{m \in \MN \where \Exists{M \in \MO(\FUS)}{\tup{m,M} \in \cH}}$.
We write $m_\cH$, where $\cH \in \FU(\FUS)$ and $m \in \IF(\cH)$, for
the unique $M \in \MO(\FUS)$ such that $\tup{m,M} \in \cH$.

We look upon the set $\IF(\cH)$, where $\cH \in \FU(\FUS)$, as the
interface of $\cH$.
It looks to be convenient to have a notation for the restriction of a
functional unit to a subset of its interface.
We write $\tup{I,\cH}$, where $\cH \in \FU(\FUS)$ and
$I \subseteq \IF(\cH)$, for the functional unit
$\set{\tup{m,M} \in \cH \where m \in I}$.

Let $\cH \in \FU(\FUS)$.
Then an \emph{extension} of $\cH$ is an $\cH' \in \FU(\FUS)$ such that
$\cH \subseteq \cH'$.

According to the definition of a functional unit,
$\emptyset \in \FU(\FUS)$.
By that we have a unique functional unit with an empty interface, which
is not very interesting in itself.
However, when considering services that behave according to functional
units, $\emptyset$ is exactly the functional unit according to which the
empty service $\emptyserv$ (the service that is not able to process any
method) behaves.

We will use \PGLBsbt\ instruction sequences to derive partial method
operations from the method operations of a functional unit.
We write $\Lf{I}$, where $I \subseteq \MN$, for the set of all \PGLBsbt\
instruction sequences, taking the set $\set{f.m \where m \in I}$ as the
set $\BInstr$ of basic instructions.

The derivation of partial method operations from the method operations
of a functional unit involves services whose processing of methods
amounts to replies and service changes according to corresponding method
operations of the functional unit concerned.
These services can be viewed as the behaviours of a machine, on which
the processing in question takes place, in its different states.
We take the set $\FU(\FUS) \x \FUS$ as the set $\Services$ of services.
We write $\cH(s)$, where $\cH \in \FU(\FUS)$ and $s \in \FUS$, for the
service $\tup{\cH,s}$.
The functions $\effect{m}$ and $\sreply{m}$ are defined as follows:
\pagebreak[2]
\begin{ldispl}
\begin{aeqns}
\effect{m}(\cH(s)) & = &
\Biggl\{
\begin{array}[c]{@{}l@{\;\;}l@{}}
\cH(m_\cH^e(s))            & \mif m \in \IF(\cH) \\
{\emptyset}(s')            & \mif m \notin \IF(\cH)\;,
\end{array}
\beqnsep
\sreply{m}(\cH(s))  & = &
\Biggl\{
\begin{array}[c]{@{}l@{\;\;}l@{}}
m_\cH^r(s) \phantom{\cH()} & \mif m \in \IF(\cH) \\
\Div                       & \mif m \notin \IF(\cH)\;,
\end{array}
\end{aeqns}
\end{ldispl}%
where $s'$ is a fixed but arbitrary state in $S$.
We assume that each $\cH(s) \in \Services$ can be denoted by a closed
term of sort $\Serv$.
In this connection, we use the following notational convention: for each
$\cH(s) \in \Services$, we write $\cterm{\cH(s)}$ for an arbitrary
closed term of sort $\Serv$ that denotes $\cH(s)$.
The ambiguity thus introduced could be obviated by decorating $\cH(s)$
wherever it stands for a closed term.
However, in this paper, it is always immediately clear from the context
whether it stands for a closed term.
Moreover, we believe that the decorations are more often than not
distracting.
Therefore, we leave it to the reader to make the decorations mentally
wherever appropriate.

Let $\cH \in \FU(\FUS)$, and let $I \subseteq \IF(\cH)$.
Then an instruction sequence $x \in \Lf{I}$ produces a partial method
operation $\moextr{x}{\cH}$ as follows:
\begin{ldispl}
\begin{aceqns}
\moextr{x}{\cH}(s) & = &
\tup{\moextr{x}{\cH}^r(s),\moextr{x}{\cH}^e(s)}
 & \mif \moextr{x}{\cH}^r(s) = \True \Or
        \moextr{x}{\cH}^r(s) = \False\;, \\
\moextr{x}{\cH}(s) & \mathrm{is} & \mathrm{undefined}
 & \mif \moextr{x}{\cH}^r(s) = \Div\;,
\end{aceqns}
\end{ldispl}%
where
\begin{ldispl}
\begin{aeqns}
\moextr{x}{\cH}^r(s) & = & x \sfreply f.\cterm{\cH(s)}\;, \\
\moextr{x}{\cH}^e(s) & = &
\mathrm{the\;unique}\; s' \in S\; \mathrm{such\;that}\;
 x \sfapply f.\cterm{\cH(s)} = f.\cterm{\cH(s')}\;.
\end{aeqns}
\end{ldispl}%
If $\moextr{x}{\cH}$ is total, then it is called a
\emph{derived method operation} of $\cH$.

\section{Functional Units for Stack Machines}
\label{sect-func-unit-sbs}

In this section, we define some notions that have a bearing on the
halting problem in the setting of \PGLBsbt\ and functional units.
The notions in question are defined in terms of functional units for the
following state space:
\begin{ldispl}
\SBS = \seqof{\set{0,1,\sep}}\;.
\end{ldispl}%

The elements of $\SBS$ can be understood as the possible contents of the
tape of a stack whose alphabet is $\set{0,1,\sep}$. It is assumed that the top
is the left-most element.

The colon serves as a separator of bit sequences.
This is for instance useful if the input of a program consists of
another program and an input to the latter program, both encoded as a
bit sequences.
We could have taken any other tape alphabet whose cardinality is greater
than one, but $\set{0,1,\sep}$ is quite handy when dealing with
issues relating to the halting problem.

Below, we will use a computable injective function
$\funct{\alpha}{\SBS}{\Nat}$ to encode the members of $\SBS$ as natural
numbers.
Because $\SBS$ is a countably infinite set, we assume that it is
understood what is a computable function from $\SBS$ to $\Nat$.
An obvious instance of a computable injective function
$\funct{\alpha}{\SBS}{\Nat}$ is the one where
$\alpha(a_1 \ldots a_n)$ is the natural number represented in the
quaternary number-system by $a_1 \ldots a_n$ if the symbols $0$, $1$, and
$\sep$ are taken as digits representing the numbers $1$,
$2$, and  $3$, respectively.

A method operation $M \in \MO(\SBS)$ is \emph{computable} if there exist
computable functions $\funct{F,G}{\Nat}{\Nat}$ such that
$M(v) = \tup{\beta(F(\alpha(v))),\alpha^{-1}(G(\alpha(v)))}$ for all
$v \in \SBS$, where $\funct{\alpha}{\SBS}{\Nat}$ is a computable
injection and $\funct{\beta}{\Nat}{\Bool}$ is inductively defined by
$\beta(0) = \True$ and $\beta(n + 1) = \False$.
A functional unit $\cH \in \FU(\SBS)$ is \emph{computable} if, for each
$\tup{m,M} \in \cH$, $M$ is computable.

It is assumed that, for each $\cH \in \FU(\SBS)$, a computable
injective function from $\Lf{\IF(\cH)}$ to $\seqof{\set{0,1}}$ with a
computable image has been given that yields, for each
$x \in \Lf{\IF(\cH)}$, an encoding of $x$ as a bit sequence.
If we consider the case where the jump lengths in jump instructions are
character strings representing the jump lengths in decimal notation and
method names are character strings, such an encoding function can
easily be obtained using the ASCII character-encoding.

Although this may be of lesser generality than possible, we will assume that
ASCII encoding is used thus removing a degree of freedom, and determining
in detail how an implementation of instruction sequence programming over
$\cH$ is supposed to work.

We use the notation $\ol{x}$ to denote the encoding of $x$ as a bit
sequence.

Let $\cH \in \FU(\SBS)$, and let $I \subseteq \IF(\cH)$.
Then:
\begin{itemize}
\item
$x \in \Lf{\IF(\cH)}$ produces a
\emph{solution of the halting problem} for $\Lf{I}$ with respect to
$\cH$ if:
\begin{ldispl}
x \cvg f.\cterm{\cH(v)}\; \mathrm{for\; all}\; v \in \SBS\;, \\
x \sfreply f.\cterm{\cH(\ol{y} \sep v)} = \True \Iff
y \cvg f.\cterm{\cH(v)}\; \mathrm{for\; all}\;
y \in \Lf{I}\; \mathrm{and}\; v \in \seqof{\set{0,1,\sep}}\;;
\end{ldispl}%
\item
$x \in \Lf{\IF(\cH)}$ produces a
\emph{reflexive solution of the halting problem} for $\Lf{I}$ with
respect to $\cH$ if $x$ produces a solution of the halting problem for
$\Lf{I}$ with respect to $\cH$ and $x \in \Lf{I}$;
\item
the halting problem for $\Lf{I}$ with respect to $\cH$ is
\emph{autosolvable} if there exists an $x \in \Lf{\IF(\cH)}$ such that
$x$ produces a reflexive solution of the halting problem for $\Lf{I}$
with respect to $\cH$;
\item
the halting problem for $\Lf{I}$ with respect to $\cH$ is
\emph{potentially autosolvable} if there exist an extension $\cH'$ of
$\cH$ and the halting problem for $\Lf{\IF(\cH')}$ with respect to
$\cH'$ is autosolvable;
\item
the halting problem for $\Lf{I}$ with respect to $\cH$ is
\emph{potentially recursively autosolvable} if there exist an extension
$\cH'$ of $\cH$ and the halting problem for $\Lf{\IF(\cH')}$ with
respect to $\cH'$ is autosolvable and $\cH'$ is computable.
\end{itemize}
These definitions make clear that each combination of an
$\cH \in \FU(\SBS)$ and an $I \subseteq \IF(\cH)$ gives rise to a
\emph{halting problem instance}.

Below we will make use of a method operation $\Dup \in \MO(\SBS)$ for duplicating
bit sequences.
This method operation is defined as follows:
\begin{ldispl}
\begin{aceqns}
\Dup(v)   & = & \tup{\True,v \sep v}
 & \mif v \in \seqof{\set{0,1}}\;, \\
\Dup(v \sep w) & = & \tup{\True,v \sep v \sep w}
 & \mif v \in \seqof{\set{0,1}}\;.
\end{aceqns}
\end{ldispl}%

\begin{proposition}
\label{prop-dup}
Let $\cH \in \FU(\SBS)$ be such that $\tup{\dup,\Dup} \in \cH$,
let $I \subseteq  \IF(\cH)$ be such that $\dup \in I$,
let $x \in \Lf{I}$, and
let $v \in \seqof{\set{0,1}}$ and $w \in \seqof{\set{0,1,\sep}}$ be such
that $w = v$ or $w = v \sep w'$ for some $w' \in \seqof{\set{0,1,\sep}}$.
Then
$(f.\dup \conc x) \sfreply f.\cterm{\cH(w)} =
 x \sfreply f.\cterm{\cH(v \sep w)}$.
\end{proposition}
\begin{proof}
This follows immediately from the definition of $\Dup$ and the axioms
for~$\sfreply$.
\qed
\end{proof}
The method operation $\Dup$ is a derived method operation of the
above-mentioned functional unit whose method operations correspond to
the basic steps that a Turing machine with tape alphabet
$\set{0,1,\sep}$ can perform on its tape.
This follows immediately from the computability of $\Dup$ and the
universality of this functional unit.

Below we
will make use of two simple transformations of \PGLBsbt\ instruction
sequences that affect only their termination behaviour on  and in particular
the Boolean value yielded at termination in the case of termination.
Here, we introduce notations for those transformations.

Let $x$ be a \PGLBsbt\ instruction sequence.
Then we write $\swap(x)$ for $x$ with each occurrence of $\haltP$
replaced by $\haltN$ and each occurrence of $\haltN$ replaced by
$\haltP$, and we write $\ftod(x)$ for $x$ with each occurrence of
$\haltN$ replaced by $\fjmp{0}$.
In the following proposition, the most important properties relating to
these transformations are stated.
\begin{proposition}
\label{prop-swap-f2d}
Let $x$ be a \PGLBsbt\ instruction sequence.
Then:
\begin{enumerate}
\item
if $x \sfreply u = \True$ then $\swap(x) \sfreply u = \False$ and
$\ftod(x) \sfreply u = \True$;
\item
if $x \sfreply u = \False$ then $\swap(x) \sfreply u = \True$ and
$\ftod(x) \sfreply u = \Div$.
\end{enumerate}
\end{proposition}
The proof is an trivial adaptation of the elementary proof of the corresponding
statement in the case of Turing Machine tapes, instead of Stack Machine data.

\section{Method names and method operations for a stack}
\label{sect-stack-methods}
At this stage it is useful to lay down the names and meaning of the common methods
for stack manipulation. This can be done in many ways, and any choice will do. The interface
$I_s$ consists of the following ten method names. These eight methods are taken together in a
functional unit $\cH_s$ that represents a stack with this particular three symbol alphabet as a
functional unit over $\SBS$.

\begin{itemize}
\item $\emptyst$ leaves the state of the functional unit unchanged and returns $\True$
if the state represent and empty stack and $\False$ otherwise.
\item $\popst$ deletes the leftmost symbol, and returns reply $\True$, if the stack is non-empty,
otherwise it leaves the stack empty and returns $\False$.
\item $\pusho$, $\pushz$ and $\pushc$ insert respectively $0, 1$ and $:$ on the left-most position
and each return $\False$.
\item $\topeqo$, $\topeqz$, and $\topeqc$ each test for the presence of a specific character at
the top of the stack.
If the stack is empty or its top differs from the symbol mentioned in the basic instruction name the reply
is $\False$, otherwise it is $\True$. In all cases the stack is left unchanged.
\end{itemize}

As mentioned above $\Dup$ is a method on stacks as well, but it is not included in the 
methods on $\cH_s$.

\section{Turing Impossibility Properties}
\label{sect-t-i-properties}
The recursive unsolvability theorem by Turing is an impossibility result which
may be found in many different circumstances. Looking at its proof
that proof establishes the negation of potential autosolvability. Subsequently
by combining it with the Church--Turing thesis that fact can be phrased in
terms of recursive solvability in general.

As an impossibility result we take Turing's theorem to establish the impossibility
of a reflexive solution of the halting problem in any functional unit in $\FU(\SBS)$
extending $\cH$. That state of affairs concerning a programming environment
will be termed the (strong) Turing impossibility property. We formulate this only for
functional units in $\FU(\SBS)$ but it should be clear that these definitions can be
adapted to many contexts that allow an encoding of programs (instruction
sequences) into the state space upon which a program is acting when executed.
Consider a functional unit $\cH \in \FU(\SBS)$, and let $I = \IF(\cH)$. The
pair $(\Lf{I},\cH)$ constitutes an instruction sequence programming
environment. For programming environments of this kind we introduce the
following notions.
\begin{itemize}
\item
The programming environment has the
\emph{strong Turing impossibility property}
if its halting problem is not potentially autosolvable. By default
Turing Impossibility refers to
strong Turing impossibility if no further qualification is provided.
\item
The programming environment has the
\emph{intermediate Turing Impossibility Property}
if its halting problem is not potentially recursively autosolvable.
\item
The programming environment has the
\emph{weak Turing impossibility property}
if its halting problem is not autosolvable.
\end{itemize}

It has been established in \cite{BM09k} and implicitly in  \cite{BP04a} that the
strong Turing impossibility property holds for some programming environments
where the halting problem is recursively solvable. This is an interesting situation
because it combines the intuitions of two seemingly incompatible
worlds: general computability on machines with
an unbounded state space where Turing impossibility is taken for granted, and the computing
devices that emerge from digitalized electrical engineering where everything is finite state and
where for that reason all problems have computable solutions, however inefficient these
solutions may be.

We have no information about the existence of programing environments that have the
intermediate Turing impossibility property but not the strong one and also not about the
existence of programming environments that satisfy the weak Turing impossibility property
and not the intermediate one. At this stage we have no indication that such examples will be of
methodological importance for the theory of computer programming.

\section{Strong Turing impossibility in the presence of $\dup$}
\label{sect-autosolvability}
The following theorem tells us essentially that potential
autosolvability of the halting problem is precluded in the presence of
the method operation $\Dup$.
\begin{theorem}
\label{theorem-non-autosolv}
Let $\cH \in \FU(\SBS)$ be such that $\tup{\dup,\Dup} \in \cH$, and
let $I \subseteq \IF(\cH)$ be such that $\dup \in I$.
Then there does not exist an $x \in \Lf{\IF(\cH)}$ such that $x$
produces a reflexive solution of the halting problem for $\Lf{I}$ with
respect to $\cH$.
\end{theorem}
\begin{proof}
Assume the contrary.
Let $x \in \Lf{\IF(\cH)}$ be such that $x$ produces a reflexive
solution of the halting problem for $\Lf{I}$ with respect to $\cH$, and
let $y = f.\dup \conc \ftod(\swap(x))$.
Then $x \cvg f.\cterm{\cH(\ol{y} \sep \ol{y})}$.
By Proposition~\ref{prop-swap-f2d}, it
follows that $\swap(x) \cvg f.\cterm{\cH(\ol{y} \sep \ol{y})}$
and either
$\swap(x) \sfreply f.\cterm{\cH(\ol{y} \sep \ol{y})} = \True$ or
$\swap(x) \sfreply f.\cterm{\cH(\ol{y} \sep \ol{y})} = \False$.

In the case where
$\swap(x) \sfreply f.\cterm{\cH(\ol{y} \sep \ol{y})} = \True$,
we have by Proposition~\ref{prop-swap-f2d} that
(i)~$\ftod(\swap(x)) \sfreply f.\cterm{\cH(\ol{y} \sep \ol{y})}
       = \True$ and
(ii)~$x \sfreply f.\cterm{\cH(\ol{y} \sep \ol{y})} = \False$.
By Proposition~\ref{prop-dup}, it follows from~(i) that
$(f.\dup \conc \ftod(\swap(x))) \sfreply f.\cterm{\cH(\ol{y})} =
 \True$.
Since $y = f.\dup \conc \ftod(\swap(x))$, we have
$y \sfreply f.\cterm{\cH(\ol{y})} = \True$.
On the other hand, because $x$ produces a reflexive solution, it follows
from~(ii) that $y \dvg f.\cterm{\cH(\ol{y})}$.
This contradicts with
$y \sfreply f.\cterm{\cH(\ol{y})} = \True$.

In the case where
$\swap(x) \sfreply f.\cterm{\cH(\ol{y} \sep \ol{y})} = \False$,
we have by Proposition~\ref{prop-swap-f2d} that
(i)~$\ftod(\swap(x)) \sfreply f.\cterm{\cH(\ol{y} \sep \ol{y})}
       = \Div$ and
(ii)~$x \sfreply f.\cterm{\cH(\ol{y} \sep \ol{y})} =
\True$.
By Proposition~\ref{prop-dup}, it follows from~(i) that
$(f.\dup \conc \ftod(\swap(x))) \sfreply f.\cterm{\cH(\ol{y})} =
 \Div$.
Since $y = f.\dup \conc \ftod(\swap(x))$, we have
$y \sfreply f.\cterm{\cH(\ol{y})} = \Div$.
On the other hand, because $x$ produces a reflexive solution, it follows
from~(ii) that $y \cvg f.\cterm{\cH(\ol{y})}$.
This contradicts with
$y \sfreply f.\cterm{\cH(\ol{y})} = \Div$.
\qed
\end{proof}
It is easy to see that Theorem~\ref{theorem-non-autosolv} goes through
for all functional units for $\SBS$ of which $\Dup$ is a derived method
operation.

Now, let $\cH = \set{\tup{\dup,\Dup}}$.
By Theorem~\ref{theorem-non-autosolv}, the halting problem for
$\Lf{\set{\dup}}$ with respect to $\cH$ is not (potentially)
autosolvable.
However, it is recursively solvable.
\begin{theorem}
\label{theorem-decidable}
Let $\cH = \set{\tup{\dup,\Dup}}$.
Then the halting problem for $\Lf{\set{\dup}}$ with respect to $\cH$ is
decidable.
\end{theorem}
\begin{proof}
Let $x \in \Lf{\set{\dup}}$, and let $x'$ be $x$ with each occurrence of
$f.\dup$ and $\ptst{f.\dup}$ replaced by $\fjmp{1}$ and each occurrence
of $\ntst{f.\dup}$ replaced by $\fjmp{2}$.
For all $v \in \SBS$, $\Dup^r(v) = \True$.
Therefore, $x \cvg f.\cH(v) \Iff x' \cvg \emptysf$ for all $v \in \SBS$.
Because $x'$ is finite, $x' \cvg \emptysf$ is decidable.
\qed
\end{proof}

\section{Open issues on Turing Impossibility properties for stack machine programming}
\label{sect-open-problems}
About Turing impossibility properties for stack machine programming we know
in fact almost nothing except the result just proven that presence of $\dup$ implies the
strong Turing impossibility property.

Let $\cH_{s}^{\prime}$ result from $\cH_{s}$ by removing the method
$\pushc$. It follows from the results in \cite{BM09k} that this functional unit yields
a programming system for which the halting problem is potentially recursively autosolvable.
The difference made by the presence of this one method is quire remarkable.

It is now easy to formulate several plausible questions which are open to the best
of our knowledge. Indeed the objective of this lengthy paper is no more than to introduce the
terminology of Turing impossibility  properties and to state
these problems in full detail. Let $\cH_{s,dup}$ denote the extension of $\cH_s$ with
the method $\dup$. $I_{s,dup}$ is its interface.

\begin{enumerate}
\item Is the halting problem for $\Lf{I_{s,dup})}$ w.r.t. $\cH_{s,dup}$ recursively solvable?%
\footnote{A simpler but  equally interesting problem results if the action $\pushc$ is removed 
from $\cH_{s,dup}$ and
from $I_{s,dup}$.}
\item If so, can  $\cH_{s,dup}$ be extended with methods that are not derivable from $\cH_{s,dup}$
 without destroying recursive solvability of the halting problem?
\item Does the programming system $\Lf{I_{s})}$ with $\cH_{s}$ feature
the weak Turing impossibility property?
\item If so, what about the intermediate and strong Turing impossibility properties?
\end{enumerate}

Some remarks concerning the motivation of there questions is in order. To begin with, the virtue
of separating Turing impossibility from recursive unsolvability is that the technical content
of the recursive unsolvability proof for the Halting problem is made independent from the
Church-Turing thesis. However convincing that thesis may be, unquestionably  it is strongly
connected with general computability theory
on the infinite set of natural numbers. As a conceptual toolkit for understanding the practice
of computation recursion theory on the natural numbers can be questioned, however.

Phrasing the halting problem in terms of program machine interaction, rather than exclusively
in terms of machines correlates with the fact that the intuition of computing on an unbounded
platform has been so successful for the development and deployment of high level program notations. Much more
so than for the area computer architecture which always keeps the underlying electric circuitry in mind,
and for which the digital perspective means that an abstraction can be made
 from infinite state machines  in need of a
probabilistic analysis to finite state machines that can be understood, at least in principle,
without the use of probabilities.

\section{Concluding Remarks}
\label{sect-concl}

We have put forward three flavors of the Turing impossibility property: strong, intermediate and weak.
These notions have been applied to stack machine programming. Some results concerning that
case have been translated from the work on Turing machines in \cite{BM09k}, and several open
questions have been formulated.

Programming environments which satisfy the strong Turing impossibility property and for which the
halting problem is recursively solvable at the same time constitute an interesting bridge between the two worlds
of computer science: general computation without bounds on memory and time, and finite state
computation in bounded time. The existence of these combined 
circumstances depends on being specific on
how the encoding of instruction sequences into data is achieved. The classical
Turing impossibility property for a Turing complete programming environment is not dependent 
on the specific way in which that encoding is done, in that sense the classical approach is more general.

\bibliographystyle{spmpsci}
\bibliography{IS}

\end{document}